\documentclass[12pt]{article} \usepackage{latexsym} \textwidth 15cm
\begin{document}

\title{{\bf DYADOSPHERES DON'T DEVELOP}
\thanks{astro-ph/0605434}}
\author{
Don N. Page
\thanks{Internet address:
don@phys.ualberta.ca}
\\
Institute for Theoretical Physics\\
Department of Physics, University of Alberta\\
Edmonton, Alberta, Canada T6G 2J1\\
and \\
Asia Pacific Center for Theoretical Physics (APCTP)\\
Hogil Kim Memorial Building \#519\\
POSTECH, San 31\\
Hyoja-dong, Namgu, Pohang\\
Gyeongbuk 790-784, Korea
}
\date{(2006 May 17)}
\maketitle
\large

\begin{abstract}

\baselineskip 18 pt

Pair production itself prevents the development of dyadospheres, hypothetical
macroscopic regions where the electric field exceeds the critical Schwinger
value.  Pair production is a self-regulating process that would discharge a
growing electric field, in the example of a hypothetical collapsing charged
stellar core, before it reached 6\% of the minimum dyadosphere value, keeping
the pair production rate more than 26 orders of magnitude below the dyadosphere
value.

\end{abstract}
\normalsize

\baselineskip 18 pt
\newpage

\section{Introduction}\label{sec1}

Ruffini and his group
\cite{DR,Ruf,PRX,RX,RSWX,BRX,RBFCX,RBFXC,RV,RXBFC,RBCFX,CMRX,RVX,RBBCFX,RBCFGX,BR}
have proposed a model for gamma ray bursts that invokes a {\it dyadosphere\/},
a macroscopic region of spacetime with rapid Schwinger pair production
\cite{Sch}, where the electric field exceeds the critical electric field value
\begin{equation} E_c \equiv {m^2\over q} \equiv {m^2c^3\over\hbar q} \approx
1.32 \times 10^{16} \:\mathrm{V/cm}. \label{eq:1} \end{equation} (Here $m$ and
$-q$ are the mass and charge of the electron, and I am using Planck units
throughout.)  The difficulty of producing these large electric fields is a
problem with this model that has not been adequately addressed.  Here I shall
summarize calculations \cite{dyad} showing that dyadospheres almost certainly
don't develop astrophysically.

The simplest reason for excluding dyadospheres is that if one had an
astrophysical object of mass $M$, radius $R > 2M$, and excess positive charge
$Q$ in the form of protons of mass $m_p$ and charge $q$ at the surface, the
electrostatic repulsion would overcome the gravitational attraction and eject
the excess protons unless $qQ \leq m_p M$ or
 \begin{equation}
 {E\over E_c} = {qQ\over m^2 R^2} \leq {m_p M\over m^2 R^2}
 < {m_p\over 4 m^2 M} < 1.2\times 10^{-13}\left({M_\odot\over M}\right),
 \label{eq:2}
 \end{equation}
where $M_\odot$ is the solar mass.  (If the excess charge were negative and in
the form of electrons, the upper limit would be smaller by $m/m_p$.)  Then the
pair production would be totally negligible.

However, one might postulate the implausible scenario in which protons are
bound to the object by nuclear forces, which in principle are strong enough to
balance the electrostatic repulsion even for dyadosphere electric fields. 
Therefore, for the sake of argument, I did a calculation \cite{dyad} of what
would happen under the highly idealized scenario in which the surface of a
positively charged stellar core with initial charge $Q_0 = M$ (the maximum
allowed before the electrostatic repulsion would exceed the gravitational
attraction on the entire core, not just on the excess protons on its surface)
freely fell from rest at radial infinity along radial geodesics in the external
Schwarzschild metric of mass $M$.

This idealization ignores the facts that a realistic charged surface would (a)
not fall from infinity, (b) have one component of outward acceleration,
relative to free fall, from the pressure gradient at the surface, (c) have
another component of outward acceleration from the electrostatic repulsion, and
(d) fall in slower in the Reissner-Nordstrom geometry if the gravitational
effects of the electric field with $Q \sim M$ were included.  Because of each
one of these effects, the actual surface would fall in slower at each radius
and hence have more time for greater discharge than in the idealized model. 
Hence the idealized model gives a conservative upper limit on the charge and
electric field at each radius, even under the implausible assumption that the
excess protons are somehow sufficiently strongly bound to the surface that they
are not electrostatically ejected.

As we shall see, even in this highly idealized model, the self-regulation of
the pair production process itself will discharge any growing electric field
well before it reaches dyadosphere values.  This occurs mainly because
astrophysical length scales are much greater than the electron Compton
wavelength, which is the scale at which the pair production becomes significant
at the critical electric field value for a dyadosphere.  Therefore, the
electric field will discharge astrophysically even when the pair production
rate is much lower than dyadosphere values.

These calculations lead to the conclusion that it is likely impossible
astrophysically to achieve, over a macroscopic region, electric field values
greater than a few percent of the minimum value for a dyadosphere, if that. 
The Schwinger pair production itself would then never exceed $10^{-26}$ times
the minimum dyadosphere value.

\section{Schwinger discharge of an electric field}\label{sec2}

In this section we shall analyze the pair production and discharge of an
electric field produced by the collapse of the idealized hypothetical charged
sphere or stellar core of mass $M$ and initial positive charge $Q_0$, assuming
that somehow the excess charge on its surface is not electrostatically ejected,
and assuming that the surface falls in as rapidly as possible, which is free
fall from rest at infinity in an assumed external Schwarzschild geometry.

As the surface radius $R$ collapses, the electric field $E=Q/r$ outside ($r>R$)
produces pairs, with the positrons escaping and the electrons propagating in to
the surface to reduce its charge $Q$.  The pair production rate $\mathcal{N}$
per 4-volume \cite{Sch} is given by
 \begin{equation}
 \mathcal{N} = {q^2 E^2\over 4\pi^3}\exp{\left(-{\pi m^2\over qE}\right)}
             \equiv {m^4\over 4\pi} {e^{-w}\over w^2},
 \label{eq:3}
 \end{equation}
where
 \begin{eqnarray}
 w &\equiv& {\pi m^2\over qE} \equiv {\pi E_c \over E} = {\pi m^2 r^2\over qQ}
 \nonumber \\
 &=&{4\pi m^2 M_\odot\over q}\left({M\over M_\odot}\right)
 \left({Q\over M}\right)^{-1}\left({r\over 2M}\right)^2
 \nonumber \\
 &\equiv&{1\over B}\left({M\over M_\odot}\right)
 \left({Q\over M}\right)^{-1}\left({r\over 2M}\right)^2,
 \label{eq:4}
 \end{eqnarray}
 \begin{equation}
 B \equiv {q\over 4\pi m^2 M_\odot} \approx 42475.
 \label{eq:5}
 \end{equation}
A dyadosphere has $E \geq E_c \equiv m^2/q \Rightarrow w \leq \pi \Rightarrow$
 \begin{equation}
 \left({M\over M_\odot}\right)
 \left({Q\over M}\right)^{-1}
 \left({r\over 2M}\right)^2
 \leq \pi B \equiv {q\over 4 m^2 M_\odot} \approx 1.33\times 10^5.
 \label{eq:6}
 \end{equation}

For astrophysical electric fields anywhere near dyadosphere values, the
electrons and positrons produced will quickly be accelerated to very near the
speed of light, so one will effectively get a null number flux 4-vector
$\mathbf{n_+}$ of highly relativistic positrons moving radially outward and
another null number flux 4-vector $\mathbf{n_-}$ of highly relativistic
electrons moving radially inward, with total current density 4-vector
$\mathbf{j} = q \mathbf{n_+} - q \mathbf{n_-}$.

It is most convenient to describe this current in terms of radial null
coordinates, say $U$ and $V$, so that the approximately Schwarzschild metric
outside the collapsing core may be written as
 \begin{equation}
 ds^2 = - e^{2\sigma} dU dV + r^2(U,V)(d\theta^2 + \sin^2{\theta} d\phi^2).
 \label{eq:7}
 \end{equation}
Then Maxwell's equations (Gauss's law) gives
 \begin{equation}
 4\pi\mathbf{j} \equiv 4\pi q(n_+^V \partial_V - n_-^U \partial_U)
 = \nabla\cdot\mathbf{F}
  \equiv {1\over r^2}\left(Q^{,V}\partial_V - Q^{,U}\partial_U \right).
 \label{eq:8}
 \end{equation}

The 4-divergence of each of the number flux 4-vectors $\mathbf{n_+}$ and
$\mathbf{n_-}$ is equal to the pair production rate $\mathcal{N}$, which leads
to the following relativistic partial differential equation for the pair
production and discharge process:
 \begin{equation}
 Q_{,UV} = -2\pi q r^2 e^{2\sigma} \mathcal{N}
  =-{q^3 Q^2 e^{2\sigma}\over 2\pi^2 r^2}
      \exp{\left(-{\pi m^2 r^2\over qQ}\right)},
 \label{eq:9}
 \end{equation}
or
 \begin{equation}
 8\pi q r^2 \mathcal{N}
  = {2q^3 r^2 E^2\over \pi^2} \exp{\left(-{\pi m^2\over qE}\right)}
  =\  ^2\!\Box (r^2E)
  = r\Box\left(rE\right) - r^3 E\Box\left({1\over r}\right),
 \label{eq:10}
 \end{equation}
where $^2\Box$ is the covariant Laplacian in the 2-dimensional metric
$^2ds^2=-e^{2\sigma}dU dV$ and where $\Box$ is the covariant Laplacian in the
full 4-dimensional metric.

In my much more detailed paper \cite{dyad}, I have analyzed this partial
differential equation for the charge distribution over the entire spacetime
region exterior to the charged surface and have found an approximation that
reduces it to a relativistic ordinary differential equation for the evolution
of the charge at the surface itself.  However, here I shall confine myself to a
Newtonian approximation to this evolution equation, which turns out to give
results very close to the relativistic approximation.

A convenient time parameter for describing the collapse of the surface is the
velocity $v$ the surface has in the frame of a static observer at fixed $r$ when
the surface radius $R$ crosses that value of $r$.  For free fall from rest at
infinity in the external Schwarzschild metric of mass $M$, with $\tau$ being the
proper time along the surface worldlines, one gets
 \begin{equation}
 v = -{dR\over d\tau} = \sqrt{2M\over R},
 \label{eq:11}
 \end{equation}
so that the proper time remaining until the proper time $\tau_c$ at which the
surface reaches the curvature singularity at $R=0$ is $\tau_c-\tau =
(4/3)M/v^3$ and $R = 2M/v^2 = (4.5 M)^{1/3}(\tau_c-\tau)^{2/3}$.  The Newtonian
limit of this is when $R\gg 2M$, which gives $v\ll 1$ and $\tau\approx t$, the
Schwarzschild coordinate time.  In this limit, the surface moves negligibly
during the time it takes for electrons to move inward from where they are
created to the surface to reduce $Q(t)$.

If $w \equiv \pi E_c/E$ is defined at each point outside the surface, let
 \begin{eqnarray}
 z(t) \equiv w(t,R(t)) \equiv {\pi E_c\over E(R(t))}
  = {\pi m^2 R(t)^2\over qQ(t)} = {\pi m^2 M^2\over q Q v^4}
 \label{eq:12}
 \end{eqnarray}
be the value of $w$ at the surface itself.  Since we shall find that the
electric field $E$ always stays far below the critical dyadosphere value $E_c$,
we have $z \gg 1$, which will be used for various approximations below.

Now the pair production rate (and assumed effectively instantaneous propagation
of the electrons produced to the surface of the collapsing stellar core) gives
 \begin{eqnarray}
 {dQ\over dt} \approx -q \int_R^\infty\mathcal{N}(r) 4\pi r^2 dr
  \approx -q{m^4R^4\over z^2} \int_R^\infty {dr\over r^2} e^{-z{r^2\over R^2}}
  \approx -{q m^4 R^3\over 2 z^{3} e^{z}}.
 \label{eq:13}
 \end{eqnarray}
Then using $R = 2M/v^2$ and $dv/dt \approx dr/d\tau = v^4/(4M)$ leads to the
following ordinary nonlinear first-order differential equation for $z(v)$ in
the Newtonian limit:
 \begin{eqnarray}
 {v\over z}{dz\over dv} \approx -4\left[1-{(Mmq)^2\over\pi v^5 z^2 e^z}\right]
 = -4\left[1-{A\mu^2\over v^5 z^2 e^z}\right]
  = -4\left[1-e^{-U}\right],
 \label{eq:14}
 \end{eqnarray}
where
 \begin{equation}
 A \equiv {(M_\odot m q)^2\over\pi} \approx 3.39643251\times 10^{28},
 \label{eq:15}
 \end{equation}
 \begin{equation}
 \mu \equiv {M\over M_\odot},
 \label{eq:16}
 \end{equation}
 \begin{equation}
 U \equiv z + \ln{z} -\ln{A} -2\ln{\mu} +5\ln{v}.
 \label{eq:17}
 \end{equation}

From Eq. (\ref{eq:12}), one can see that the boundary conditions for Eq.
(\ref{eq:14}) are that initially ($\tau\approx t=-\infty \Leftrightarrow
R=\infty \Leftrightarrow v=0$) $v^4 z = \pi m^2 M^2/(qQ)$ with the surface
charge $Q$ having its asymptotically constant initial value $Q_0$, which will
be taken to be its maximum allowed value, $M$, unless otherwise specified.  One
can see from this that both $z$ and $U$ start off initially at infinite
values.  The final value will be when the surface enters the event horizon of
the black hole at $R=2M$ or $v=1$.  This is beyond the applicability of the
Newtonian approximation being used here, but it turned out that the
relativistic analysis \cite{dyad} gave very nearly the same answers.

One can now differentiate Eq. (\ref{eq:17}) for $U(v)$, using Eq. (\ref{eq:14}),
to obtain
 \begin{equation}
 v{dU\over dv} \approx -4(z+2)\left(1-e^{-U}\right) + 5.
 \label{eq:18}
 \end{equation}
Since $z\gg 1$, this equation implies that $U$ decreases to near zero (though
it cannot reach zero, for if it could, the right hand side would be positive,
contradicting the assumption that it dropped to zero and hence had a negative
derivative on the left hand side).  Then when $U \ll 1$, Eq. (\ref{eq:17}) may
be solved approximately for $z(v)$ to give
 \begin{eqnarray}
 z(v) &\approx& \ln{A}+2\ln{\mu}-5\ln{v}-2\ln{(\ln{A}+2\ln{\mu}-5\ln{v})}
  \nonumber \\
  &\approx& \ln{A} -2\ln{\ln{A}} +
  \left(1-{2\over\ln{A}}\right)(2\ln{\mu}-5\ln{v})
  \nonumber \\
  &\approx& 57.33 + 1.94 \ln{\mu} + 4.85 \ln{1\over v}.
 \label{eq:19}
 \end{eqnarray}
This then gives the ratio of the electric field at the surface to the critical
electric field of a dyadosphere as being
 \begin{equation}
 {E\over E_c} = {\pi\over z}
  \approx 0.0548-0.00185\ln{\mu}-0.00463\ln{1\over v} \ll 1.
 \label{eq:20}
 \end{equation}

One can improve this result by using a slightly improved formula (giving a
roughly 2.5\% correction for $z \sim 57$) for the radial integral in Eq.
(\ref{eq:13}) for the pair production rate, by using a better explicit
approximation for what $U$ should approach at $v=1$, and by using a numerical
solution of the thus-corrected form of Eq. (\ref{eq:17}) for $z$ with this
expression for $U$, to estimate that when $\mu=1$ and $v=1$, $z \approx
57.5843$.  Including the improved formula for the right hand side of Eq.
(\ref{eq:13}) into the Newtonian differential equation (\ref{eq:14}) and then
integrating it numerically from $v=0$ to $v=1$ for $\mu=1$ gave the result at
the horizon of the solar black hole of $z \approx 57.5845$, so only the 6th
digit changed from the algebraic estimate obtained without numerically solving
the Newtonian differential equation.

In \cite{dyad} I used a relativistic ordinary differential equation
approximation to the partial differential equation (\ref{eq:9}) and was able to
deduce an explicit approximate relativistic result for $\mu=1$ and $v=1$ of $z
\approx 57.60483$, whereas the numerical solution of the relativistic ordinary
differential equation gave $z \approx 57.60480$, differing from the explicit
formula (using as input the values of $m$, $q$, and $M_\odot$) by only about
one part in two million.  However, I would estimate that the relativistic
ordinary differential equation approximation to the partial differential
equation would itself introduce absolute errors of the order of
$10^{-4}$--$10^{-3}$ in the value of $z$, so the numerical solution of that
ordinary differential equation is not necessarily any better than the
completely explicit approximate solution I also obtained.

The difference between the Newtonian and the relativistic approximations for
$z$ on the horizon ($v=1$) of a solar mass collapsing core ($\mu=M/M_\odot=1$)
is about 0.02, which I would guess is considerably larger than the error of my
relativistic approximation (not given here, but in \cite{dyad}), but it is
still a relative difference of only about one part in three thousand for the
idealized upper limit of the value of the electric field of a hypothetical
charged stellar core collapsing into a black hole after falling in freely (no
nongravitational forces on the surface) from starting at rest at radial
infinity with the external metric being Schwarzschild.

We can also give a heuristic derivation of the Newtonian result in the following
way:  We expect the self-regulation of the electric field to make $z = \pi E_c/E
= \pi m^2 R^2/(qQ)$ change slowly, so
 \begin{equation}
 {1\over Q}{dQ\over dt} \sim {2\over R}{dR\over dt}
  \approx -{1\over M} \left({2M\over R}\right)^{3\over 2} = -{v^3\over M}.
 \label{eq:21}
 \end{equation}
Then the pair production rate per 4-volume, $\mathcal{N} = (m^4/4\pi)e^{-w}/w^2
\approx (m^4/4\pi z^2)e^{-zr^2/R^2}$ for $z\gg 1$ decreases roughly
exponentially with $r$ with e-folding length $\Delta r \sim R/(2z)$.  Thus
 \begin{eqnarray}
 {1\over Q}{dQ\over dt}
  &\approx& -{q\over Q} \int_R^\infty\mathcal{N}(r) 4\pi r^2 dr
  \approx -{q\over Q}\mathcal{N}(r=R) 4\pi R^2 \Delta r
  \nonumber \\
  &\approx& -q{qz\over\pi m^2 R^2}{m^4\over 4\pi z^2}e^{-z}4\pi R^2{R\over 2z}
  = -{1\over M}{M^2 m^2 q^2\over\pi}{e^{-z}\over z^2}\left({R\over 2M}\right).
 \label{eq:22}
 \end{eqnarray}
Equating this to $-(1/M)(2M/R)^{3/2}$ gives
 \begin{equation}
 1 \approx {M^2 m^2 q^2\over\pi}{e^{-z}\over z^2}
 \left({R\over 2M}\right)^{5\over 2} = {A\mu^2\over v^5}{e^{-z}\over z^2},
 \label{eq:23}
 \end{equation}
which implies $z+2\ln{z} \approx \ln{A} + 2\ln{\mu} -5\ln{v}$, just as we got
from the approximated solution of the ordinary differential equation in the
Newtonian approximation.

A dyadosphere would have $E \geq E_c$, which implies $z=\pi E_c/E \leq z_c =
\pi$ and $\mathcal{N} = (m^4/4\pi)e^{-z}/z^2 \geq \mathcal{N}_c =
m^4 e^{-\pi}/(4\pi^3)$.  But
 \begin{eqnarray}
 \mathcal{N} &\approx& {m^4\over 4\pi}{v^5\over A\mu^2}
 = {\pi^2 e^{\pi} v^5\over A\mu^2}\mathcal{N}_c
 = {\pi^3 e^{\pi} v^5\over q^2 m^2 M} \mathcal{N}_c
 \nonumber \\
 &\approx& 0.672 \times 10^{-26} \, {v^5\over\mu^2}\mathcal{N}_c
  < 10^{-26} \, \mathcal{N}_c,
 \label{eq:24}
 \end{eqnarray}
so the heuristic estimate gives the maximum pair production rate more than 26
orders of magnitude below that of a dyadosphere.  By comparison, the numerical
solution of the approximate relativistic ordinary differential equation
\cite{dyad} gave, at $v=1$,
 \begin{eqnarray}
 {\mathcal{N}\over \mathcal{N}_c} \approx {0.661168\times 10^{-26}\over \mu^2}
    (1 + 0.0005545  \ln{\mu} - 0.00001759 \ln^2{\mu}),
 \label{eq:25}
 \end{eqnarray}
about 1.6\% less than the heuristic estimate gives.

One can also calculate the maximum efficiency for converting the collapsing
stellar core mass $M$ into outgoing positron energy,
 \begin{eqnarray}
 \epsilon &\approx& - \int{QdQ\over MR}
  \approx {2\mu^2\over B^2} \int_0^1 e^{-U} {dv\over v^7 z^2}
  \nonumber \\
  &\approx& {1\over 3}B^{-1/2}
  \left[\ln{A}+{5\over 4}\ln{B}-{3\over 4}\ln{\ln{(AB^2)}}
        +{3\over 4}\ln{\mu}\right]^{-1/2}\mu^{1/2}(Q_0/M)^{3/2}
  \nonumber \\
  &=&{m\over 3M}\sqrt{4\pi Q_0^3\over q}
  \left[{1\over 4}\ln{\left({q^{13}M^3\over 2^{10}\pi^9 m^3}\right)}
  -{3\over 4}\ln{\ln{\left({q^4\over 16\pi^3 m^2}\right)}}\right]^{-1/2}
   \nonumber \\
   &\sim& 1.9\times 10^{-4}\left[1+{\ln{(M/M_\odot)}\over 300}\right]^{-1/2}
   \left({M\over M_\odot}\right)^{1/2} \left({Q_0\over M}\right)^{3/2}
   \nonumber \\
   &\sim& 2\times 10^{-4}\left({Q_0\over M}\right)^{3/2}\sqrt{M\over M_\odot}.
 \label{eq:26}
 \end{eqnarray}

By comparison, the numerical solution of the approximate relativistic ordinary
differential equation \cite{dyad} gave the more precise result
 \begin{equation}
 \epsilon \approx 0.0001855 \left({M\over M_\odot}\right)^{0.495}
  \left({Q_0\over M}\right)^{0.742}.
 \label{eq:27}
 \end{equation}
The dominant factor in an estimate of the coefficient 0.0001855 (the upper limit
on the efficiency if $Q_0=M=M_\odot$) is one-third the ratio of the electron
mass to the proton mass, which is 0.0001815.  This efficiency is too low for the
pair production from these idealized collapsing charged cores to explain gamma
ray bursts, even if it is admitted that this very implausible scenario (of the
excess charge $Q \sim M$ not getting expelled from the collapsing core by the
huge electrostatic forces on it) comes nowhere near being able to form a
dyadosphere.

One can also calculate \cite{dyad} that the probability of one of the particles
annihilating with an antiparticle is less than $10^{-26}$, so the direct
interactions of individual particles is negligible, consistent with what was
assumed above.

\section{Conclusions}

If protons are bound to a collapsing stellar core purely gravitationally, the
maximum electric field is more than 13 orders of magnitude below dyadosphere
values:  $E_{\mathrm{max}} \leq 1.2\times 10^{-13} (M_\odot/M) E_c$.

If protons are much more strongly bound, $E_{\mathrm{max}} \leq 0.055 E_c$ and
$\mathcal{N}_{\mathrm{max}} \leq 10^{-26} \mathcal{N}_c$, where $\mathcal{N}_c$
is the minimal dyadosphere production rate.

The energy efficiency of the process for $M \sim M_\odot$ is very low,
$\epsilon \stackrel{<}{\sim} 1.86\times 10^{-4}(M/M_\odot)^{1/2}(Q_0/M)^{3/2} \approx
(Q_0/M)^{3/2}\sqrt{M/(2.9\times 10^7 M_\odot)}$.

If one relaxed the assumptions of this model, such as the spherical symmetry,
one would expect to get similar results, perhaps changing the pair production
rates by factors of order unity that depend upon the precise geometry. 
However, it seems very unlikely that any modification could increase the
maximum possible pair production rate by any significant fraction of the 26
orders of magnitude that the model above fails to achieve dyadosphere values. 
Therefore, the example analyzed strongly suggests that dyadospheres do not form
astrophysically.

\section*{Acknowledgments}

I am grateful for being able to participate in the 9th Italian-Korean Symposium
on Relativistic Astrophysics in South and North Korea and there learn about
dyadospheres.  I am thankful for the hospitality of Sang Pyo Kim at Kunsan
University and for the hospitality of Beijing Normal University at the
beginning of this work.  Financial support was also provided in part by the
Natural Sciences and Engineering Research Council of Canada



\baselineskip 6pt

\begin{thebibliography}{99}

\bibitem{DR} T. Damour and R. Ruffini, Phys. Rev. Lett. {\bf 35}, 463-466
(1975).

\bibitem{Ruf} R. Ruffini, ``Beyond the Critical Mass: The Dyadosphere of
Black Holes,'' \emph{Black Holes and High Energy Astrophysics, Proceedings of
the Yamada Conference XLIX on Black Holes and High Energy Astrophysics held on
6-10 April, 1998 in Kyoto, Japan}, eds. H. Sato and N. Sugiyama (Frontiers
Science Series No. 23, Universal Academic Press, Tokyo, 1998), p.167; ``On the
dyadosphere of black holes,'' astro-ph/9811232; Astronomy and Astrophysics
Supplement {\bf 138}, 513-514 (1999).

\bibitem{PRX} G. Preparata, R. Ruffini, and S.-S. Xue, Astron. Astrophys. {\bf
338}, L87-L90 (1998), astro-ph/9810182; Nuovo Cim. {\bf B115}, 915 (2000); J.
Korean Phys. Soc. {\bf 42}, S99-S104 (2003), astro-ph/0204080.

\bibitem{RX} R. Ruffini and S.-S. Xue, ``Radiation and Time Scale Evolution of
a P.E.M. Pulse from EMBH,'' \emph{Abstracts of the 19th Texas Symposium on
Relativistic Astrophysics and Cosmology, held in Paris, France, Dec. 14-18,
1998}, eds. J. Paul, T. Montmerle, and E. Aubourg (CEA Saclay, 1998).

\bibitem{RSWX} R. Ruffini, J. Salmonson, J. Wilson, and S.-S. Xue, Astronomy
and Astrophysics Supplement {\bf 138}, 511-512 (1999), astro-ph/9905021;
Astronomy and Astrophysics {\bf 350}, 334-343 (1999), astro-ph/9907030;
Astronomy and Astrophysics {\bf 359}, 855-864 (2000), astro-ph/0004257.

\bibitem{BRX} C. L. Bianco, R. Ruffini, and S.-S. Xue, Astronomy and
Astrophysics {\bf 368}, 377-390 (2001), astro-ph/0102060.

\bibitem{RBFCX} R. Ruffini, C. L. Bianco, F. Fraschetti, P. Chardonnet, and
S.-S. Xue, Nuovo Cim. {\bf B116}, 99 (2001), astro-ph/0106535.

\bibitem{RBFXC} R. Ruffini, C. L. Bianco, F. Fraschetti, S.-S. Xue, and P.
Chardonnet, Ap. J. {\bf 555}, L107-L111 (2001), astro-ph/0106531; L113-L116
(2001), astro-ph/0106532; L117-L120 (2001), astro-ph/0106534.

\bibitem{RV} R. Ruffini and L. Vitagliano, Phys. Lett. {\bf B545}, 233-237
(2002), astro-ph/0209072.

\bibitem{RXBFC} R. Ruffini, S.-S. Xue, C. L. Bianco, F. Fraschetti, and P.
Chardonnet, La Recherche {\bf 353}, 30-32 (2002).

\bibitem{RBCFX} R. Ruffini, C. L. Bianco, P. Chardonnet, F. Fraschetti, and
S.-S. Xue, Ap. J. {\bf 581}, L19-L22 (2002), astro-ph/0210648; Int. J. Mod.
Phys. {\bf D12}, 173-270 (2003), astro-ph/0302141.

\bibitem{CMRX} P. Chardonnet, A. Mattei, R. Ruffini, and S.-S. Xue, Nuovo
Cim. {\bf 118B}, 1063-1070 (2003).

\bibitem{RVX} R. Ruffini, L. Vitagliano, and S.-S. Xue, Phys. Lett. {\bf B559},
12-19 (2003), astro-ph/0302549; Phys. Lett. {\bf B573}, 33-38 (2003),
astro-ph/0309022.

\bibitem{RBBCFX} R. Ruffini, M. G. Bernardini, C. L. Bianco, P. Chardonnet, F.
Fraschetti, and S.-S. Xue, Advances in Space Research {\bf 34}, 2715-2722
(2004), astro-ph/0503268.

\bibitem{RBCFGX} R. Ruffini, C. L. Bianco, P. Chardonnet, F. Fraschetti, V.
Gurzadyan, and S.-S. Xue, Int. J. Mod. Phys. {\bf D13}, 843-852 (2004),
astro-ph/0405284.

\bibitem{BR} C. L. Bianco and R. Ruffini, Ap. J. {\bf 605}, L1-L4 (2004),
astro-ph/0403379; {\bf 620}, L23-L26 (2005), astro-ph/0501390; {\bf 633},
L13-L16 (2005), astro-ph/0509621.

\bibitem{Sch} F. Sauter, Z.\ Phys.\  {\bf 69}, 742 (1931); W.~Heisenberg and
H.~Euler, Z.\ Phys.\  {\bf 98}, 714 (1936); V.~Weisskopf, K.\ Dan.\ Vidensk.\
Selsk.\ Mat.\ Fys.\ Medd.\ {\bf 14}, No. 6 (1936); J.~S.~Schwinger, Phys.\
Rev.\  {\bf 82}, 664 (1951); A.~I.~Nikishov, Nucl.\ Phys.\ B {\bf 21}, 346
(1970).

\bibitem{dyad} D. N. Page, ``No Astrophysical Dyadospheres,'' astro-ph/0605432.

\end{thebibliography}
\end{document}